# NEUTRINO ASTRONOMY WITH HIGH SPATIAL RESOLUTION IS ALREADY EXISTING

**V.A. Rantsev-Kartinov**[ξ]

*RRC "Kurchatov Institute", Kurchatov Sq. 1, 123182, Moscow, Russia*

*Abstract*

By basing on observations of skeletal structures of the Sun and assuming that some of them are located inside of star, and also that a filamentary (linear) matter (whose a model earlier was put forward by B.U. Rodionov) is in basis of these internal structures the author consider possible processes of images formation of these structures inside the Sun and theirs coming out into space and also gives an elementary estimations of its parameters, which allow: i) to form their images in a flux of electronic neutrinos; ii) to carry out these images from within of the Sun into space; iii) to develop these images in form of a concomitant flux of soft x-ray, which next is recorded by telescope of soft x-ray. It is supposed the processes considered here, actually, can be accepted as future base of neutrino astronomy with high spatial resolution.

## I. INTRODUCTION

The analysis (by means of the method of multilevel dynamical contrasting (MMDC) developed and described by the author earlier [1a, 1b]) of images of the Sun in range of waves lengths of a soft x-ray has resulted in revealing of fractal skeletal structures (FSS) of the Sun [2]. The topology of these structures appeared identical to the same that was revealed and described earlier in a wide range of spatial scales, the phenomena and environments [3]. MMDC analysis of images of solar spots (SS) in soft x-ray has shown they have three-dimensional structure of the same topology (see Fig. 1).

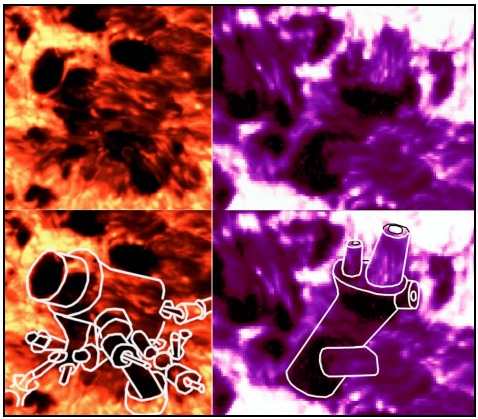

**Figure 1.** The two fragments of image of active zone of the Sun surface and below its schematic presentations are presented here. It is visible the SS presents coaxial tubular structures of the Sun of filamentary matter which are pushed out from bowels of our star to its surface. The topology of these structures is the same that in observable before in dust deposit of tokomak T-10. Tridimentionality of the structures, their inter-lacings and connections are precisely traced here. Diameters of these structures are ~ 2 $10^9$ cm and their upper butt-ends are towered above the Sun surface up to height ~ $10^{10}$ cm.

Hence from this in [2] the hypothesis has been put forward - *the Sun has internal FSS which can be by development of one of possible forms of filamentary matter (FM)* (for example, which was suggested by B.U.Rodionov [4]), *and SS are part of FSS of the Sun which have been squeezed out during its activity from within to its surface*. The revealed fragments of FSS of the Sun which have radius of rotation around of a star axis less radius of its disk on breadth of their locations (see Fig. 2) can be good acknowledgement of this hypothesis.

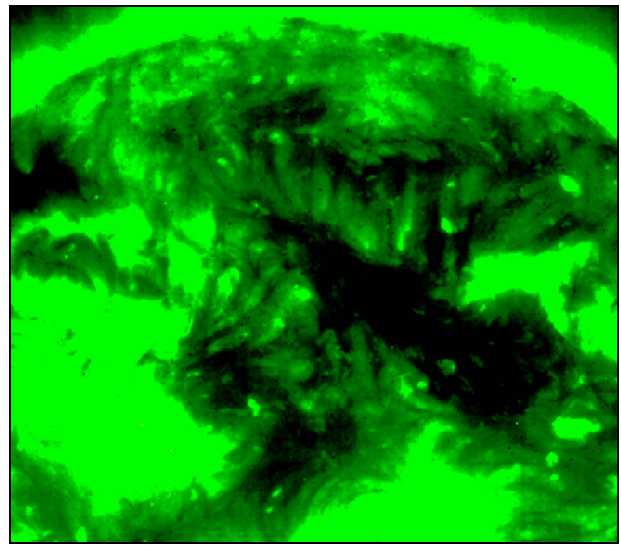

**Figure 2.** Fragment of the Sun image, = 195 Å (SOHO). The edge of a solar disk is seen overhead. Complexly weaved structure of the FM, which has coaxial tubular structures (CTS) collected of similar blocks but of the smaller size is visible. Diameter of CTS ~ 3.3 $10^{10}$ cm, di-

---

[ξ] email:rank@nfi.kiae.ru

ameters of blocks (of which they are collected) are ~ $10^9$ cm, diameter of the FM (going out from the center of its butt-end) is ~ $8 \cdot 10^9$ cm. This CTS has radius of its rotation around of the Sun axis of some smaller than radius of solar disk on breadth of its location.

As soft x-ray cannot take out the image of "interior" of the Sun into a space the author considers the following general scheme of its formation: 1) neutrinos by passing through FSS of the Sun FM are coding and take out into space their image in their own flux; 2) this image transforms by any mechanism into an accompanying flux of x-ray quantums which develops on the screen of a telescope. For realization of such scheme it is necessary to describe the prospective nature and properties of FM of the Sun and to find mechanisms of formation and developing of its image in x-ray.

## II. PARAMETERS OF MODIFIED MODEL OF FILAMENTARY MATTER

Suggested the FM model not up to the end still is developed and described, but it is suitable for primary estimations of basic parameters of such matter. Filaments of such FM represents infinite chains of quarks in the Bose-condensate state (BCS) strung onto quantified a magnetic field flux which minimal value is equal $\varphi_0 = $ hc/e. Here, h - the Plank constant, c - velocity of light in vacuum, e - a charge of electron. Let's consider that all quarks are in BCS, they are relativistic and have identical energy ~ 5 MeV. Then the quark filament radius ($r_q \sim 5 \cdot 10^{-13}$ cm) is defined by size of the de Broil wave, a magnetic field intensity (flux of which into one quantum connects everything quarks of such infinite filament into one whole) inside of such filament is $H_q \sim 3 \cdot 10^{17}$ Gs and energy of quarks connection is ~ 5 GeV. According to the model, such quark filament can be surrounded by an electron cloud in BCS. The minimal radius of such shell of relativistic electrons, which is covering a back magnetic flux ($\varphi_0$) is $r_e \sim 3 \cdot 10^{-12}$ cm, and a field inside this shell is $H_e \sim 7 \cdot 10^{15}$ Gs.

The most important result of the analysis of revealed structures in the universe (with participation of the author) is its fractality. Above described model of FM does not allow building structures of generations. Therefore the author modified it, assuming, the FM was formed at very hot universe. The quark filaments of final length in those days could have a shell (intercepting the back flux of a magnetic field) too of quarks. Internal cords and external shells of extended filaments of such matter can present consecutive chains of quarks in BCS with various combinations of charge composition. In particular, it can exist filaments which are consisting of identical sequence of the quarks which are in BCS ($2 \times (-1/3) + 2 \times (+2/3) + 2 \times (-1/3) + \cdot \cdot$). The charge agreeing with unit of length of any-one filament is equal to zero. Extended filaments of FM can be assembled of separate cylindrical blocks the length of which is connected with constant thin structure $\alpha = 1/137$. These blocks of the minimal length can be accepted as "linear atoms" (LA) of such FM. From here for estimated calculations it is possible to believe, that quantity of nucleons in such LA of FM $N_N \sim 6 \cdot 10^2$, linear density of nucleons in it $n_N \sim 6 \cdot 10^{13}$ cm$^{-1}$, its mass $m_{aq} \sim 1.2 \cdot 10^{-21}$ g, length $l_a \sim 10^{-11}$ cm, the relation of its length to diameter (which it is equal ~ $4r_q \sim 2 \cdot 10^{-12}$ cm) ~ 5. The density of substance of FM is ~ $4 \cdot 10^{13}$ cm$^{-3}$, i.e. only in 5 times there is less than density of a nuclear matter. The LA of FM can build FSS, [3], showing their universality. As average density of substance of Sun $\rho_C \sim 10$ g/cm$^3$ the volume fraction of FM inside a star should be small. The LA radius of FM in $10^4$ times is less than radius of hydrogen atom it is neutral and does not interact with usual atoms. Interaction of this matter with usual nuclei must be extremely weak. Such FM can exist neutral and inside the Sun as the temperature even of its most bowels is much lower than energy of connection even of electrons in external shell of its filaments inasmuch as energy of connection of electrons in FM is ~ 5 МэВ. For such LA a linear density of electrons in them is $n_e \sim 3 \cdot 10^{13}$ cm$^{-1}$, mass $m_{ae} \sim 6 \cdot 10^{-22}$ g ~ $3 \cdot 10^2$ $m_p$, where $m_p$ - mass of a proton, the linear density of substance in them ~ $6 \cdot 10^{-11}$ g·cm$^{-1}$, the relation of length to its diameter is ~ 2.

## III. FRACTAL SKELETAL STRUCTURES OF FILAMENTARY MATTER

Inasmuch as the topology of revealed FSS of the Sun was found identical to one which was before observable in the dust carbon deposits taken from chambers of plasma installations, the model of FSS construction of the FM of LA has been chosen similar to construction FSS of carbon nanotubes [3]. The mass, number of electrons, length and diameter of tubes of generations with number n is in this case completely determined and the latest can be estimated along formulas (1):

$$d_n \sim 6 \cdot 10^{-12} \cdot 5^{n-1}\ cm.;\ l_n \sim 2d_n \sim 2 \cdot 6 \cdot 10^{-12} \cdot 5^{n-1}\ cm., \quad (1)$$

From here blocks of various generations of such FSS have scale factor K = 5. As tubular filament of 2-nd generations with radial connections consists of ~ 80 LA, that number of LA in such filament of generation with number n is $N_l^n \sim (80)^{n-1} = 2^{3(n-1)} \cdot 10^{n-1}$, full number of electrons in it is $N_{le}^n \sim N_N \cdot N_l^n = 3 \cdot 10^2 \cdot 2^{3(n-1)} \cdot 10^{n-1} = 3 \cdot 10^2 \cdot 2^{3(n-1)} \cdot 10^{n+1}$, its volume is $V_l^n \sim 0.25 \cdot \pi \cdot d_n^2 \sim 4.3 \cdot 10^{-34} \cdot 5^{3(n-1)}$ cm$^3$ and average electrons density in it is $n_{le}^n \sim N_{le}^n / V_l^n \approx 2^{4n+32}/5^{2n-38}$ cm$^{-3}$. Observable filaments of FM in images of the Sun have the characteristic size in diameter ~ $3 \cdot 10^9$ cm. From here, according to expression (1), we finds number of generation of blocks of this structure, n ~ 30 and then according to above we calculates average density of electrons in these filaments, $n_{le}^n \sim 2.5 \cdot 10^{30}$ cm$^{-3}$.

## IV. FORMATION AND DEVELOPMENT OF IMAGES OF THE FRACTAL SKELETAL STRUCTURES OF FILAMENTARY MATTER OF THE SUN

Average density of the Sun matter is $\sim 10 \cdot$g$\cdot$cm$^{-3}$, average electrons density of its plasmas is $n_e^C \sim 5 \cdot 10^{24}$ cm$^{-3}$ $\sim 5 \cdot 10^5 \cdot n_{le}^n$ cm$^{-3}$. On sharp gradients of electron density, occurs amplified oscillations of electronic neutrinos and turning of their some part into muonic ones [5], it is especial if a condition of a resonance for this process is satisfied. Thus, the image of FSS of the Sun FM can form in a neutrino flux, which is coming through a body of the Sun in a direction of the observer, inasmuch in areas with rather homogeneous electron density neutrino do not cooperate almost with substance. For those regions for which the condition of the resonant oscillations is satisfied the image contrast in the neutrinos flow can make up value almost 100 %. Neutrino dispersion on filaments of FM, due to a high tension of a magnetic field in them, can promote also for formation of FSS image. Thus, the neutrino flow is able to forming and freely to carry over the image of internal structure of the star from within into outside. ]

Further there is a problem connected with displaying of such image. The fact is that we do not have the corresponding screen, capable to show such image. Because of very small value of interaction cross-section, even thickness of sphere of the Earth is not capable to play a role of such screen. Here the author puts forward a hypothesis, that in space between the Earth and the Sun there are same structures of FM, but rarer as they belong to considerably higher numbers of generations. The electronic neutrinos passing across fibers of FM, and cooperating with their magnetic field (according to conclusions of paper [6]) are able to generate the quantums extending in a direction of the observer.

It is necessary to note here, that only at first sight, the setting of problem about of electromagnetic interactions of neutrinos can appear absurd inasmuch initially Pauli postulated neutrino as electrically neutral particles, i.e., as a noninteracting particles with electromagnetic fields. However it turned out, that electromagnetic properties at the massive neutrino arise at the account of its interaction with vacuum of Standard model. According to theory Weinberg – Glashow - Salam, the neutrino which is moving inside an external electromagnetic field, at the time moment of t in a point with coordinate **r** with some probability breaks up onto a virtual electron and $W^+$ - бозон, and at the time moment of t′ in a point with coordinate **r'** virtual the electron and $W^+$ - бозон mutually are absorbing each other, and converting again into the real neutrino. Quantum-mechanical indeterminacy principle, «time - energy», $\Delta E \Delta t \geq \hbar/2$ allows existence of such particles during of small time intervals of the order $\Delta t \sim \hbar/\Delta E$. For the virtual $W^+$ - boson the estimation of this time gives: $\Delta t \sim \dfrac{\hbar}{m_W c^2} \approx 2 \cdot 10^{-27} c.$

Inasmuch a virtual particles taking place in the given process are a charged particles, then their interaction with an external magnetic field changes their state. In such case we have amendments to movement of the massive neutrinos. This amendment can be expressed as the amendment to value of the neutrino mass provided that the impulse neutrino does not vary: $\Delta E = (mc^4/E)\Delta m$. One of amendments to energy can be the energy of interaction of magnetic moment of neutrino, $\mu_\nu$, with an external magnetic field, $H$, ($U_H = -(\mu_\nu H)$). Differently, the Dirac massive neutrino at interaction with vacuum obtains a magnetic moment. At that it is accepted definition: the magnetic moment of neutrino is directed lengthways of neutrino spin, and for antineutrino - against it.

At calculations of radiation amendments to the neutrino mass it is possible to take account influence of an external electromagnetic field precisely. It is allowing revealing a dynamic nature of magnetic moment of the massive neutrino and mass. It turned out, that they are complex nonlinear functions of $H$ and $E$. If electromagnetic fields are weak, i.e., $E, H << B_0$, where $B_0 = \dfrac{m_e^2 c^3}{e\hbar} = 4.4 \cdot 10^{13}$ Gs, (here $m_e$ - electronic mass) that the magnetic moment of Dirac neutrino (in frames of the Standard model) accepts its the static value which (in system of units where $\hbar = c = 1$) is equal [6]:

$$\mu_\nu^0 = \dfrac{3eG_F m_\nu}{8\pi^2 \sqrt{2}} \approx 3 \cdot 10^{-19} \mu_0 \dfrac{m_\nu}{1 \, эB} \approx A \cdot \dfrac{m_\nu}{1 \, эB} \text{ eV/Gs} \qquad (2).$$

Here $G_F = 1.03 \cdot 10^{-5} m_p^{-2}$ - constant of Fermi $m_p$ - mass of proton, $\mu_0 = \dfrac{e\hbar}{2m_e c} = 5.8 \cdot 10^{-9}$ eV/Gs – electronic magneton of Bor, A=1.74 10$^{-27}$. From here, it follows - $\mu_\nu^0 \rightarrow 0$, at $m_\nu \rightarrow 0$. According to [6], for a case of a weak enough of the constant magnetic field ($H << B_0 \lambda$, where $\lambda = \left(\dfrac{m_W}{m_e}\right)^2 \approx 10^{11}$) and in presence of large value of a transverse impulse of neutrino ($p_\perp >> m_W c$) the neutrino magnetic moment demonstrates dependence, as from intensity of a magnetic field, so from energy of neutrino. The theoretical substantiation of this is given in [6]. For our case it is possible to consider $\mu_\nu \approx \mu_\nu^0$.

At linear along field - ($\mu_\nu^0 H$) of approximation [6], and homogeneous magnetic field - ($\mu_\nu^0 H$), the energy neutrino - ($E_\nu^H$) will be written down as:

$$E_\nu^H = E_\nu \left(1 - \zeta \mu_\nu^0 H \frac{E_{\nu\perp}}{E_\nu^2}\right), \qquad (3)$$

where:

$$E_\nu^2 = m_\nu^2 + p_\nu^2, \quad E_{\nu\perp}^2 = m_\nu^2 + p_{\nu\perp}^2, \qquad (4)$$

and spinal a number $\zeta = \pm 1$ sets orientation of particle spin lengthways or against direction of magnetic field. From laws of conservation of energy and impulse if photons are emitted it is following, the frequency of emitted photons are determined by expression (5):

$$\omega = \begin{cases} 2 \dfrac{\mu_\nu^0 H}{\hbar} \dfrac{(1-\beta_z^2)^{1/2}}{1-\beta\cos\Omega} & \text{at } \zeta = -1,\ \zeta' = +1, \\ 0 & \text{in all other cases} \end{cases} \qquad (5)$$

where, $\beta_z = v_z/c$ – longitudinal (in relation to direction of field) component of velocity neutrino, $\Omega$ – corner between neutrino impulse $\vec{p}$ and photon $\vec{k}$. From (5) it is visible, only such neutrino can emit photon the spin of which is directed against magnetic field ($\zeta = -1$), and radiation is accompanied by change of projection value onto direction of magnetic field: $\zeta = -1 \to \zeta = +1$. For construction of image which was formed in the neutrino flux (by means of a flux of quantums obtained from them) it is necessary to put $\Omega = 0$, i.e., these quantums should have the same direction of spreading, as neutrinos. Precisely the same formula has been obtained for radiation of neutron moving across of constant magnetic field earlier [7], theirs identity is evident if instead of magnetic moment of neutrino $\mu_\nu^0$, to take magnetic moment of neutron $\mu_n$.

Now, on the basis of obtained image in a range of lengths of waves of soft x-ray (100-400) Å, if energy of solar neutrinos is $E_\nu \sim 0.4$ (MeV), we will able to estimate, what rest mass of neutrino should be in order to the formula (5) was satisfied. This waves range corresponds to interval of frequencies $\omega \sim (2-0,5) \cdot 10^{17}$ s$^{-1}$.

Let's consider a case of perpendicular distribution of neutrino flux concerning direction of magnetic field, and $\omega \sim 10^{17}$ s$^{-1}$ which suits to experimental observations. Now, on the basis of expressions (2) and (5), for the given case it is possible to write down expression which gives the top limit for definition of rest mass of neutrino:

$$m_\nu \leq \frac{4AE_\nu^2 H}{\hbar\omega} \sim 1.7 \cdot 10^{-17} H \ (eV). \qquad (6)$$

In papers [6] and [7] expressions of probability of quantums radiation of light by the neutron and neutrino at movement of theirs in the constant magnetic field are listed, accordingly. On the basis of the expressions which are given in these papers, it is possible to show, that the full probability of quantums radiation at movement neutrino across a direction of homogeneous magnetic field ($H$) with taking into account (2) and (6) is proportional to $H^8$ a value:

$$W_{\nu\perp} \propto \frac{(\mu_\nu^0)^5 H^3}{1-\beta^2} \propto \frac{H^8}{1-\beta^2}, \qquad (7)$$

i.e., quickly grows with increase of the magnetic field value. This value quickly grows with increase of power of homogeneous magnetic field, and also at $\beta \to 1$. From the analysis of expression (7) it follows, that the estimation of probability of the given process needs to be carried out or along fields for which scale of lengths of radiated waves have maximal power of the magnetic field or when $\beta \to 1$. Here, it is needed to consider a three cases when the flux of the Dirac neutrino generates quantums at its interaction with: **a)** a dipole of magnetic field of the Sun (which at the highest activity can reach value $\sim 10^2$ Gs; **b)** with a magnetic flux of base filaments of FM, which are taking place in cosmos between the Earth and the Sun and connected with fractal structure of FM inside a star (i.e. located near to a surface of a star); **c)** with a magnetic flux of base filaments of FM, which is taking place in space between the Earth and the Sun and connected with fractal structure of FM of the Earth (i.e. taking place near to a surface of the Earth). Inasmuch near to the Sun surface, a power of magnetic dipole field, density of basic filaments of FM and flux neutrino have the greatest values therefore the basic contribution to formation of soft x-ray quantums (which have predominant direction - to the Earth) will give areas closely adjoining to the Sun surface. Expression (6) for our case **(a)** gives $m_\nu^a \leq 1.8 \cdot 10^{-15}$ eV, and in the case **(b)** - $m_\nu^b \leq 5.1$ eV, that practically coincides with value obtained in experiments. As the neutrino magnetic moment is proportional to its mass (see (2)) and very small, that (according to (7)) the probability of process for the case **(a)** (although, $1-\beta^2 \sim 2 \cdot 10^{-41}$) is very small value too. Estimations of probability of emission of quantums by the neutrino flux for a case **(c)** are showing the probability of emission of quantums in this case in 300 times is less than in a case **(b)**. So, absolutely clearly, the basic contribution to quantums radiation by flux neutrinos can give born quantums only in the case **(b)**. At the same time there is no necessity in averaging of magnetic field over volume of superficial layer occupied with FM struc-ture of such generation a diameter of which is little biger than the Sun disk diameter. However and in this case our estimations have shown, that at taking into account of described above model of construction of skeletal structures of FM and gathering of quantums during of several tens of minutes by entrance surface of telescope ~ $10^4$ cm$^2$), it is impossible to obtain dense enough flux of attendant quantums (which are generated due to interaction of flux solar neutrinos with magnetic field of FM located near to the Sun surface) for construction of image of solar insides.

If neutrinos are driving in very strong and variable magnetic field they also are able to radiate attendant quantums. Inasmuch as LA of FM inside itself carry direct and back the magnetic fluxes (which are quantized), the neutrino movement across of FM with oriented blocks actually means their movement in strongly variable magnetic field which frequency of change is determined by average of linear density of LA of FM along a trajectory. The layer of such oriented LAs of FM can be formed of free LAs which are not included in rotating FSS of the Sun which form original halo on the distance determined by equality of gravitation and centrifugal force acting on them. The estimation gives such halo will be on distance ~ $10^8$ km from a star. As length of wave in which have been obtained images of the Sun (and which were discussed here) is ~ $10^{-6}$ cm, then the average linear density of LAs of FM in halo should be ~ $10^6$ cm$^{-1}$. As the possible mechanism of development of the latent image in the neutrino flux, it would be possible to consider also their interaction with probable halo from particles of "a dark matter» (DM), but for this purpose it is necessary to know, first of all, its physical nature which remains a riddle for us in present time, and its interaction with FSS of FM and neutrino. Here it is necessary to note onto that fact, the LAs of FM may be considered also as the DM particles. If the conclusion of the author about observation by him of FSS of FM appears precisely confirmed, then we know the answer of a problem about the mechanism of conversion of the image of interiors of the Sun coded in a neutrino flux, and therefore it will be necessary to find this mechanism only. At the given moment the estimations, obtained from the analysis of a database of images of the Sun shows that at the taking into account of the model of construction of FSS of FM described above and gathering of quantums during several tens of minutes by the area of an entrance lens of a telescope by square ~ $10^4$ cm$^2$, the obtained flux of accompanying x-ray quantums (generated for the account not for a while yet of the unknown mechanism) is found by sufficient for development of the image of insides of the Sun.

At the same time the spatial resolution can reach value up to 5 $10^7$ cm that almost in $10^4$ times higher than it was been obtained in up-to-the-minute the Superkamioknade project [8] (see item 5). More detailed consideration of the given problem is complicated because an all available calculations of probability of radiation of quantums are carried out for movement neutrino in a homogeneous magnetic field while evidently, the neutrino pass across fibers of FM inside of which there is high power of magnetic field that can lead to essential increase of probability of radiation and which can take place here.

## V. SUPERKAMIOKNADE NEUTRINO TELESCOPE

Abundantly clearly, the neutrino telescope can be created and on the basis of direct neutrino detecting. It has been carried out at experiments SUPERKAMIOKANDE (Japan). The principle of action of this telescope is based on the fact the vigorous neutrino, passing through environment and cooperating with electrons gives to them the significant impulses which is directed along a neutrino trajectory. Such electrons emit a cone of light of Cerenkov radiation in a direction of their movement. This radiation is registered by specially created sensitive photomultipliers. The cylindrical tank filled highly with cleared water is the base of a telescope. Its height is 36 m and diameter - 34 m. The full weight of a tank is 50 kiloton. Walls and a bottom of this tank are covered with assembly with 11146 specially designed and very sensitive photomultipliers with low-noise photo-cathodes in diameter 50 cm. The arrangement of these photomultipliers was such that all sites of volume of this tank were looked through by them with big reliability. Due to creation of this telescope the image of the Sun almost for 3 years of an exposition has been obtained. The remarkable result of this huge, laborious work is submitted in a Fig. 3.

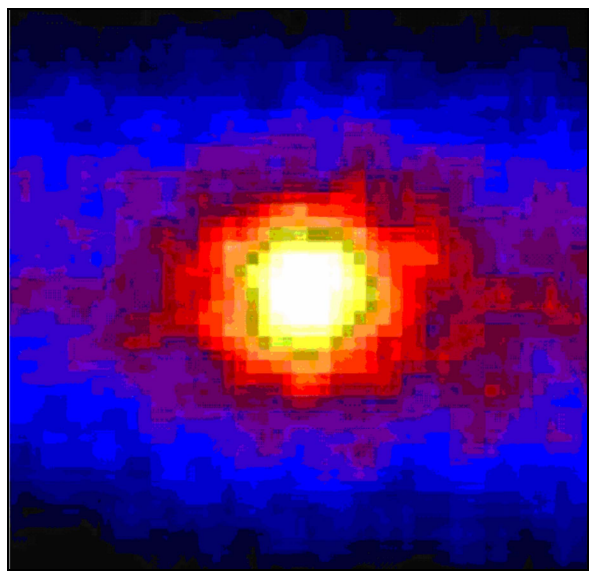

**Figure 3.** This is a neutrino profile of the Sun which was been obtained with help Super-Kamiokande of the detector [8]. One pixel of the image corresponds to one degree which should be compared with a half of degree of a solar seen disk. Distribution of contrast is determined by a corner of neutrino dispersion on electrons.

## VI. NEUTRINO ASTRONOMY OF THE HIGH SPATIAL RESOLUTION

The modern model of the Sun guarantees thermonuclear reactions only in its central part. Direct detecting of a solar neutrino flux should correspond to scale of the image of one third of its visual disk. Modern the neutrino telescope, [8], is not capable to register inside spatial structures even of the nearest star, especially of far galaxies. The size of the image of a solar disk (see Fig. 3), obtained with its help for 2.5 years in a neutrino flux, appeared almost in 25 times more of visual one and 75 times more the size corresponding to neutrino-active area of the star.

How it was been shown above if solar neutrino are Dirac-neutrino then they possess by not the zero magnetic moment and at interaction with a magnetic field they are able to radiate quantums of light. Inasmuch as value of magnetic moment of the neutrino is very small, in order the probability of such process was not zero, very big intensity of a magnetic field is necessary. It can take place only inside filaments of the matter which was been considered above. If suggested FM is a reality then there can be and an neutrino astronomy of the high spatial resolution. The fact is that the image of "insides" of the star is coded in a neutrino flux given birth inside of the same star. Further, this flux carry out image to outside and (by means of interaction with FSS of FM of the Sun and halo of oriented, free LAs or DM particles) translates this image into coded (but already in a flux of attendant quantums) the image, decoding of which occurs on the screen of a telescope. Presence of halo can give an explanation to occurrence of a horizontal strip in the image of a solar disk in a neutrino flux (see Fig. 3), because of neutrino dispersion on it. The vertical width of this strip, with taking into account of geometrical optics, corresponds to distance of this halo from the observer.

Construction of images in an optical or x-ray range has very high spatial resolution. Therefore the described above neutrino astronomy can take place only under condition of existence of FM, similar which has been described in item 2. The examples presented in paper shows that the hypotheses suggested by the author can have a reality and are the facts of development of FSS of FM inside the Sun and its nearest space environment.

## VI. SUMMARY

The fact of observation of the images displaying internal structure of the star, can solve many problems connected to neutrino physics: a) to prove existence of neutrino oscillations, i.e., that solar neutrinos are Dirac-neutrinos; b) to study of FSS of FM and their dynamics inside stars/galaxies through the analysis of their images in various ranges of lengths of waves; c) to study dynamic character of the moment and neutrino mass and their interactions with LA of FM or DM particles.

Thus, already now we have a neutrino astronomy which allows us to look into bowels of stars and galaxies, to observe their internal structure and to study processes taking place inside them, to reveal FSS of FM and to study their properties, as inside stars, as in their environment. All this can give a new push into researching of space objects, understanding of processes of their formation, and searching a new energy sources taking place in the universe.

## VII. ACKNOWLEDGMENTS

The author is deeply grateful to V.I. Kogan for invariable support and interest to these researches.

## VIII. REFERENCES

1. A.B.Kukushkin, V.A.Rantsev-Kartinov, "Self-similarity of plasma networking in a broad range of length scales: from laboratory to cosmic plasmas", RSI, vol. 70, pp. 1387-1391, (1999).
2. Rantsev-Kartinov V.A., "Revelation of the Sun Self-Similarity Skeletal Structures", in Proc. 32nd EPS Conference on Plasma Phys. Tarragona, 27 June - 1 July 2005 ECA Vol.29C, P-2.155 (2005), http://eps2005.ciemat.es/papers/pdf/P2_155.pdf.
3. A.B.Kukushkin, V.A.Rantsev-Kartinov: a) "Similarity of skeletal objects in the range $10^{-5}$ cm to $10^{23}$ cm", Phys.Lett. A, vol. 306, pp. 175-183, (2002); b) "Skeletal structures in high-current electric discharges and laser-produced plasmas: observations and hypotheses", Ed. F. Gerard, Nova Science Publishers, New York, Advances in Plasma Phys. Research, vol. 2, pp. 1-22, (2002).
4. Rodionov B.U., "Thready (linear) dark matter possible displays", Gravitation and Cosmology, vol. **8**, Supplement I, pp. 214-215, (2002).
5. Gonzalez-Garcia M.C., C.N. Yang and Yosef Nir, "Neutrino Masses and Mixing: Evidence and Implications", Rev. Mod. Phys., vol. 75, p. 345, (2003).
6. Borisov A.V., Zukovskiy V.Ch., Ternov A.I., "Electromagnetic properties of the Dirac massive neutrino at an external electromagnetic field", Izvestiya vuzov Ser. Fiz. 3, pp. 64-70, (1988), (in russian).
7. Borisov A.V., Vshivcev A.S., Zukovskiy V.Ch., Eminov P.A., Uspekhi Fizicheskikh Nauk, 167, № 3, (1997) 241-267, (in russian).
8. McDonald, Art; Spiering, Christian; Schönert, Stefan; Kearns, Edward T.; Kajita, Takaaki, "Astrophysical Neutrino Telescopes", RSI, vol. 75, № 2, p. 293, (2004).